\documentclass{article}
\usepackage{graphicx,epsfig}
\usepackage{latexsym}
\usepackage{amsfonts,amsbsy,amssymb,amsmath}
\usepackage{color}

\begin{document}

\begin{center}

\noindent {\Large \bf Interacting Constituents in Cosmology}

\vskip 1.0cm \noindent { R. Aldrovandi, R. R. Cuzinatto and L. G.
Medeiros\footnote{\, To whom all correspondence should be addressed.
E-mail: leo@ift.unesp.br.}} \vskip 0.5cm {\it Instituto de
F\'{\i}sica Te\'orica, Universidade Estadual Paulista
\\ Rua Pamplona 145, 01405-900 S\~ao Paulo SP, Brazil.}
\date{\today} \vskip 1.5cm

\end{center}

\begin{abstract}
\noindent {\small{

Universe evolution, as described by Friedmann's equations, is
determined by source terms fixed by the choice of pressure $\times$
energy-density equations of state $p\left( \rho \right) $. The usual
approach in Cosmology considers equations of state accounting only
for kinematic terms, ignoring the contribution from the interactions
between the particles constituting the source fluid. In this work
the importance of these neglected terms is emphasized. A systematic
method, based on the Statistical Mechanics of real fluids, is
proposed to include them. A toy-model is presented which shows how
such interaction terms can engender significant cosmological
effects.

}}

\vskip 1.0cm \noindent {\bf Keywords}: {Cosmology, Statistical
Mechanics, Equation of State} \\ \noindent {\bf PACS}: {98.80.-k,
05.20.-y, 05.30.-d, 95.30.Tg}

\end{abstract}

\vskip 1.5cm


\section{Introduction}

The greatest achievement of Physical Cosmology is, up to now, the Big Bang
standard model. This model is based on two principles which are consistent
with large--scale observations: the \textit{cosmological principle} and the
\textit{universal time principle}. The first postulates that the universe
space-section $\Gamma $ is homogeneous and isotropic; the latter states that
the topology of the space-time four-dimensional manifold $\mathcal{M}$ is
the direct product\ $\mathcal{M}=\mathbb{R}\times \Gamma $, and results from
the adoption of a cosmological time as parameter of the manifold foliation
\cite{nar, DiracFolh}. These principles lead~ \cite{FL,RW} to the
Friedmann-Lema\^{\i}tre-Robertson-Walker (FLRW) interval\footnote{%
\thinspace\ We shall be using natural units, $c=1,\hbar =1$.}
\begin{equation}
ds^{2}=dt^{2}-a^{2}(t)\left[ \frac{dr^{2}}{1-kr^{2}}+r^{2}d\theta
^{2}+r^{2}sen^{2}\theta d\phi ^{2}\right] \text{\ \ \ }\left( k=0,\pm
1\right) ~.  \label{zero1}
\end{equation}%
On the other hand, the source terms are described by the perfect fluid
energy-momentum tensor. This tensor $T_{\mu \nu }$ and the above interval,
when substituted into the Einstein's equations of General Relativity
\begin{equation}
R_{\mu \nu }-{\textstyle\frac{1}{2}}R\ g_{\mu \nu }-\Lambda g_{\mu \nu }=%
\frac{8\pi G}{c^{4}}\ T_{\mu \nu }\ ,  \label{cosmoEinstein}
\end{equation}%
(with a cosmological--constant $\Lambda $--term) lead to the Friedmann
equations for the scale factor $a\left( t\right) $:
\begin{eqnarray}
\frac{\dot{a}^{2}}{a^{2}} &=&\frac{\Lambda }{3}+\frac{8\pi G}{3}\rho -\frac{k%
}{a^{2}}~,  \label{quinze a dot} \\
\frac{\ddot{a}}{a} &=&\frac{\Lambda }{3}-\frac{4\pi G}{3}\left( \rho
+3p\right) ~.  \label{quinze a ddot}
\end{eqnarray}%
The solution for $a\left( t\right) $\ is obtained after inserting in these
equations the constitutive relation between the pressure $p$\ and the
energy-density $\rho $\ of the cosmic source fluid.

The relation between $p$ and $\rho $ is given directly by an \textit{%
equation of state} (EOS) of type $p=p\left( \rho \right) $, or
indirectly via the \textit{distribution functions}. Either way, here
are implicitly in use the methods of Statistical Mechanics (SM),
which can, for pedagogical purposes, be divided into two branches
\cite{balescu}: equilibrium SM, with its ensemble technique and its
partition functions; and non-equilibrium SM, whose distribution
function entail, for example, kinetic equations as the Boltzmann and
the Vlassov-Landau equations. The distribution function are not only
useful to describe systems out of equilibrium, but also systems in
thermodynamical equilibrium, to which a temperature can be
attributed. The ensemble approach, however, can only be applied to
systems in equilibrium.

Cosmology uses both sections of Statistical Mechanics, e.g.: the EOS
obtained from a partition function leads to the description of the
radiation--dominated era in the thermal history of the universe and
leads also to the $\Lambda $CDM model for the present-day cosmic
dynamics \cite{wein,Solutions}; the distribution function\
determined from Boltzmann equations are used in the perturbative
cosmological models describing the cosmic microwave background
anisotropies, as well as the formation of structures~\cite{Dod}. But
the fact is that, perturbative or not, \textbf{the present
formulation of Cosmology does not take into account the interactions
between the constituents as direct sources of gravitation}. Only
kinematical terms are computed in the ordinary (non-perturbed)
cosmological models, interactions being just used to explain
pair-production and thermalization; the dynamical terms --which
could be introduced via particle-to-particle potentials or through
the S-matrix -- are simply forgotten. Even the perturbative
approach~\ \cite{Dod} considers the interaction so insufficiently
that the pressure function appearing in the expression of the
energy-momentum tensor\ presents kinematical terms solely
\cite{Ber}.

In section \ref{sec - falta}, we shall make explicit the absence of
dynamical terms (interaction terms) both in the conventional cosmological
models developed under the hypothesis of thermodynamical equilibrium -- with
their typical EOS -- and in the perturbative models (out of equilibrium) --
built on the distribution functions. Based on the Statistical Mechanics of systems with interactions \cite%
{balescu,Pat} we discuss, in section \ref{sec - inclusao}, how to include
those terms in Cosmology. Section \ref{sec - exemplo} presents a toy-model
which shows the decisive influence that interaction terms can have on cosmic
evolution. The conclusions are summed up in section \ref{sec - Dis. Final}.

\section{Absence of interaction terms \label{sec - falta}}

This section deals, in the context of the standard model, with an ideal
source cosmic fluid, interactions between components being supposed absent.
We leave to the next section the discussion about the real need for the
inclusion of interaction terms in some phases of the universe evolution.

First, it is important to establish the period during which
equilibrium Statistical Mechanics is applicable. This is restricted
by the fact that the universe, described by the FRW metric
(\ref{zero1}), is growing up~---~the volume containing the cosmic
fluid is expanding. In the primeval universe ($kT>20$ MeV, where $k$
is the Boltzmann constant), the typical reaction rates $\Gamma
_{pri}$\ involving the different constituents are much larger than
the expansion rate
$H_{pri}$ \cite{kolb}, i.e.%
\begin{equation}
\Gamma _{pri}\gg H_{pri}\equiv \frac{\dot{a}_{pri}}{a_{pri}}~.  \label{zero2}
\end{equation}%
For the energy values then prevalent the thermodynamic notion of \textit{%
quasi-static expansion} holds: for each infinitesimal variation of volume,
all constituents of the fluid are at the same temperature~---~they keep
themselves in thermal equilibrium. Notice in advance that, as the curvature
is negligible $\left( k=0\right) $\footnote{%
\thinspace\ In agreement with the recent observational data, we will assume $%
k=0$ all along the text.} in the early universe, the ordinary
Statistical Mechanics defined on
euclidean 3-space $\mathbf{E}^{3}$ can be used. We emphasize that Eq.(\ref%
{zero2}) is valid even during an accelerated expansion.

The components of the cosmic fluid will decouple progressively as the
temperature decreases \cite{kolb,Dod}. A natural question turns up: once
decoupled from the other fluid components, will a given particular component
remain in thermal equilibrium~? This question is better formulated in terms
of the component distribution function: is there an equilibrium distribution
function for the decoupled component valid in general in an expanding plane
space~? The answer is no \cite{SchuSpi}. The proof of this statement is not
trivial, but it is related to the fact that we do not have spatially
constant time-like Killing vectors in plane FRW space-times \cite{Ber}.

Nevertheless, it is possible to construct distribution functions for the
expanding space in particular cases, as in the presence of non-relativistic
and ultrarelativistic components. This is comforting enough since, from the
early to the present-day universe, the decoupling stable particles are
either ultrarelativistic -- photons and neutrinos -- or non-relativistic
(baryonic and dark matter).\footnote{%
\, We adopt cold (non-relativistic) dark matter from the begining.} We are
therefore allowed to use equilibrium SM when working with this content. The
non-equilibrium treatment will be required only for perturbative
approximations.

Next sub-section deals specifically with the description given by the
cosmological models in thermodynamical equilibrium, and \ref{sec-OutEquil}\
discuss the models with components out of equilibrium.


\subsection{Cosmology for systems in equilibrium\label{sec-Equil}}

The classical Cosmology text-books, e.g. \cite{nar,wein,Dod,kolb}, teach us
how to determine the pressure $p_{i}$, the energy density $\rho _{i}$ and
the numerical density $n_{i}$ of the $i$-th component of the fluid in
thermal equilibrium:
\begin{subequations}
\begin{gather}
n_{i}=\frac{g_{i}}{(2\pi )^{3}}\int f_{i}(\vec{p})d^{3}p,  \label{um} \\
\rho _{i}=\frac{g_{i}}{(2\pi )^{3}}\int E_{i}(\vec{p})f_{i}(\vec{p})d^{3}p,
\label{dois} \\
p_{i}=\frac{g_{i}}{(2\pi )^{3}}\int \frac{\left\vert \vec{p}\right\vert ^{2}%
}{3E_{i}(\vec{p})}f_{i}(\vec{p})d^{3}p,  \label{tres}
\end{gather}%
where $g_{i}$ is the degeneracy degree; $E_{i}$, the dispersion relation

\end{subequations}
\begin{equation}
E_{i}^{2}=\vec{p}{\,}^{2}+m_{i}^{2}~;  \label{quatro}
\end{equation}%
and $f_{i}(\vec{p})$ the distribution functions given by%
\begin{equation}
f_{i}(\vec{p})=\frac{1}{e^{\beta (E_{i}-\mu _{i})}\pm 1 }~,\text{ \ \ \ \ \
\ \ }\beta \equiv \frac{1}{kT}~.  \label{cinco}
\end{equation}%
The lower (upper) sign refers to the Fermi-Dirac (Bose-Einstein) statistics.

The sum over all the $i$ components is taken into the Friedmann equations (%
\ref{quinze a dot}, \ref{quinze a ddot}). Standard texts on Statistical
Mechanics \cite{Pat} derive $p_{i}$, $\rho _{i}$\ and $n_{i}$\ in the
ensemble formalism for \textit{ideal} relativistic quantum particles. The
grand-canonical partition function $\Xi _{i}$ or the potential $\Omega _{i}$
for the $i$-th component are given as
\begin{eqnarray}
\Omega _{i}(V,\beta ,\mu _{i}) &\equiv &\frac{\ln \Xi _{i}(V,\beta ,\mu )}{V}%
=\pm \,\frac{g_{i}}{(2\pi )^{3}}\int \ln \left[ 1\pm z_{i}e^{-\beta E_{i}}%
\right] d^{3}p=  \notag \\
&=&\frac{g}{(2\pi )^{3}}\sum\limits_{j=1}^{\infty }\frac{(\mp 1)^{j-1}}{j}%
\int z_{i}e^{-j\beta E_{i}}d^{3}p~,  \label{seis}
\end{eqnarray}%
where
\begin{equation}
z_{i}=e^{\beta \mu _{i}}=e^{\beta \left( \mu _{i}^{NR}+m_{i}\right)
} \label{z}
\end{equation}%
is the fugacity. In (\ref{z}), the chemical potential $\mu _{i}=\mu
_{i}^{NR}+m_{i}$ has a non-relativistic contribution $\mu _{i}^{NR}$, which
is the usual term appearing in text-books. We have above included also the
term $m_{i}$\ associated to the rest-energy of the particle under
consideration. The quantities $p_{i}$, $\rho _{i}$ and $n_{i}$ are, then,
\begin{subequations}
\begin{gather}
p_{i}=\frac{1}{\beta }\Omega _{i}~;  \label{sete} \\
n_{i}=z_{i}\left. \frac{\partial \Omega _{i}}{\partial z_{i}}\right\vert
_{V,~\beta }~;  \label{oito} \\
\rho _{i}=-\left. \frac{\partial \Omega _{i}}{\partial \beta }\right\vert
_{V,~z_{i}}~.  \label{nove}
\end{gather}%
These prescriptions are equivalent to the definitions (\ref{um}), (\ref{dois}%
) and (\ref{tres}).

Analyzing the ultra-relativistic limit, $kT\gg m$, in Eqs.(\ref{dois}, \ref%
{tres}) and assuming $kT\gg \mu $ (as is the case for the photons), we
obtain
\end{subequations}
\begin{equation}
p_{\gamma }=\frac{\rho _{\gamma }}{3}\,\, ,  \label{dez}
\end{equation}%
the familiar equation of state for radiation.

Conversely, if we take the non-relativistic limit $kT\ll m$ in the same Eqs.(%
\ref{dois}, \ref{tres}) and neglect quantum effects, it results
\begin{subequations}
\begin{gather}
\rho _{NR}=n_{NR}m~,  \label{onze} \\
p_{NR}=n_{NR}kT~.  \label{doze}
\end{gather}%
The quantum effects will be negligible if the condition $n\lambda ^{3}\ll 1$
is satisfied, $\lambda $ being the thermal wavelength
\end{subequations}
\begin{equation}
\lambda =\sqrt{\frac{2\pi }{mkT}}~.  \label{thermal wavelength}
\end{equation}%
This will hold when no particle invades any other's effective volume, of
which a rough estimate is $\lambda^3 $: this is the meaning of the condition
above.

Pressure $p$ only appears in the Friedmann equations added to $\rho $. If we
consider $kT\ll m$, Eqs.(\ref{onze}, \ref{doze}) will say that%
\begin{equation}
p_{M}=0  \label{treze}
\end{equation}%
is a very good approximation for the equation of state for non-relativistic
matter $M$ (the dust approximation).

The Friedmann equations, together with (\ref{dez}) and (\ref{treze}), enable
us to calculate simplified, but analytic, solutions for the universe
evolution determined by $a\left( t\right) $ \cite{Solutions}. Present-day
universe, for example, is well described by the solution obtained\ after
inserting EOS (\ref{treze}) in (\ref{quinze a dot}, \ref{quinze a ddot}),
keeping $\Lambda $ non-null, and neglecting the contribution of
radiation~---~the so-called $\Lambda $CDM model. For the early universe,
period during which radiation dominates, the suitable EOS is (\ref{dez}).

Anyway, neither the EOS simpler forms (\ref{dez}, \ref{treze}) nor the
complete expressions, Eqs.(\ref{sete}, \ref{oito}, \ref{nove}), take
interactions between the constituents into account. Partition function (\ref%
{seis}) includes all the possible states of the $i$-th component in an \emph{%
ideal} gas, of \emph{non-interacting}\, fermionic (or bosonic) relativistic
particles. The first term of series (\ref{seis}) refers to the classical
description (Boltzmann statistics), while the others are  order-by-order
quantum corrections. We shall see in section \ref{sec - inclusao} that the
expression of the partition function in terms of a series suggests a
mechanism of inclusion of the interactions order by order.

We remind the reader that our interest here is to analyze EOS based on first
principles; Eqs.(\ref{dez}) and (\ref{doze}) are examples of this class. We
shall exclude any EOS used in the context of Cosmology constructed on
phenomenological basis, as EOS for scalar fields~\cite%
{CalDaveStein,Stein,Zlatev,FraRos,PeRa}, EOS for the Chaplygin gas~\cite%
{Fabris et al,Alca Jain, Avelino et al,Amendola et al,Bento et al} and Van
der Walls' EOS~\cite{Italia}, to mention but a few.


\subsection{Cosmology for systems out of equilibrium\label{sec-OutEquil}}

As said before, the use of non-equilibrium Statistical Mechanics is
essentially required only in perturbative cosmology. According to \cite%
{Dod,Ber}, perturbations on the FRW metric come from interactions between
the components of the cosmic fluid as described by the Boltzmann equation,
\begin{equation}
\frac{df_{i}}{dt}=\sum\limits_{j}C_{ij}[f_{i}]~,  \label{dezesseis}
\end{equation}%
where $f_{i}$\ is the distribution function of the $i$-th component and $%
C_{ij}[f_{i}]$ is the collision term for the $(i, j)$ pair of components. It
is written as

\begin{equation}
C_{ij}[f_{i}(\vec{p}_{1})]=\sum\limits_{\vec{p}_{2},~\vec{p}_{3},~\vec{p}%
_{4}}\left\vert M_{ij}\right\vert ^{2}~\left[ f_{i}(\vec{p}_{3})f_{j}(\vec{p}%
_{4})-f_{i}(\vec{p}_{1})f_{j}(\vec{p}_{2})\right] ~,  \label{dezessete}
\end{equation}%
where $M_{ij}$ is the scattering amplitude of two interacting particles with
incoming momenta $\vec{p}_{1}$ and $\vec{p}_{2} $ and out-coming momenta $%
\vec{p}_{3}$ and $\vec{p}_{4} $.

Once the distribution function is evaluated (via Boltzmann equation) for all
components of the fluid, the energy-momentum tensor is obtained as
\begin{equation}
T^{\mu \nu }=\sum\limits_{i}g_{i}\int f_{i}(\vec{p})\,\frac{p^{\mu }p^{\nu }%
}{p_{0}}\frac{d^{3}p}{(2\pi )^{3}}~,  \label{dezoito}
\end{equation}%
where $p^{\mu }$ is the comoving momentum encapsulating, of course, the
energy and the tri-momentum $\vec{p}$,
\begin{equation}
p_{0}=E~;\;\;\;\;\;\;\;\;\;p^{2}=g_{ij}p^{i}p^{j}~.  \label{dezoito
e meio}
\end{equation}%
As sources in Einstein's equations, $T^{00}$ is the energy density (\ref%
{dois}) and $T^{ii}$\ is the pressure (\ref{tres}). At first sight
one could think that the Boltzmann equation (\ref{dezesseis})
introduces interaction terms as direct sources of gravitation, but
in fact the usual pressure expression only partially considers
dynamical (interaction)
effects.  Interactions are taken  into account  only indirectly, through deformations in the distribution functions  $f_{i}$ and $%
f_{j}$. Nevertheless, the distributions  used are those of free
particles. An example of dynamical pressure appears in the van der
Walls equation
\begin{equation}
p=\frac{nkT}{(1-An)}-B \, n^{2}  \label{dezenove},
\end{equation}%
in which $A$ and $B$  \cite%
{Ber} are constants  phenomenologically fitted  for each gas. It
cannot  be obtained from (\ref{dezoito}).

Let us see in more detail what happens. The deduction of the Boltzmann
equation uses the fact that the number of pairs of particles with velocities
$\vec{v}_{i}$ and $\vec{v}_{j}$ at time $t$ is%
\begin{equation}
f_{i}(\vec{x}_{i},\vec{v}_{i})f_{j}(\vec{x}_{j},\vec{v}_{j})~.  \label{vinte}
\end{equation}%
On the other hand, Statistical Mechanics tells us that this is true only
when the correlations between particles are negligible. In fact, the number
of pairs of particles with velocities $\vec{v}_{i}$ and $\vec{v}_{j}$ is
determined by the two-particle distribution function
\begin{equation}
f_{ij}(\vec{x}_{i},\vec{x}_{j},\vec{v}_{i},\vec{v}_{j})  \label{vinte um}
\end{equation}%
and, in general,
\begin{equation}
f_{ij}(\vec{x}_{i},\vec{x}_{j},\vec{v}_{i},\vec{v}_{j})\neq f_{i}(\vec{x}%
_{i},\vec{v}_{i})f_{j}(\vec{x}_{j},\vec{v}_{j})~.  \label{vinte
dois}
\end{equation}

The $N$-particle distribution function $f_{N}(x_{1},...,x_{N})$\ may be
expressed as \cite{balescu}%
\begin{equation}
f_{N}(\vec{x}_{1},...,\vec{x}_{N})=\prod\limits_{i}^{N}f_{i}(\vec{x}_{i})+%
\bar{g}_{N}(\vec{x}_{1},...,\vec{x}_{N})~;  \label{vinte tres}
\end{equation}%
$\bar{g}_{N}$ determines the complete correlation degree of the system. For
practical reasons, it is convenient to split the set of $N$ particles in all
the possible disjoint subsets containing at least one particle, i.e.%
\begin{eqnarray}
f_{i}(\vec{x}_{i}) &=&f_{i}(\vec{x}_{i})~,  \notag \\
f_{ij}(\vec{x}_{i},\vec{x}_{j}) &=&f_{i}(\vec{x}_{i})f_{j}(\vec{x}%
_{j})+g_{ij}(\vec{x}_{i},\vec{x}_{j})~,  \notag \\
f_{ijk}(\vec{x}_{i},\vec{x}_{j},\vec{x}_{k}) &=&f_{i}(\vec{x}_{i})f_{j}(\vec{%
x}_{j})f_{k}(\vec{x}_{k})+f_{i}(\vec{x}_{i})g_{jk}(\vec{x}_{j},\vec{x}_{k})+
\label{vinte seis} \\
&&+f_{j}(\vec{x}_{j})g_{ik}(\vec{x}_{i},\vec{x}_{k})+f_{k}(\vec{x}%
_{k})g_{ij}(\vec{x}_{i},\vec{x}_{j})+g_{ijk}(\vec{x}_{i},\vec{x}_{j},\vec{x}%
_{k})~,  \notag
\end{eqnarray}%
etc. In these equations $f_{i}$ is the one-particle distribution function;\ $%
f_{ij}$ is the two-particle distribution function, and so on.

This is the \textit{cluster} expansion formalism, function $g_{s}(\vec{x}%
_{1},...,\vec{x}_{s})$\ being the irreductible correlation function of $s$
particles. These are the functions describing interactions. For example, the
van der Walls equation can be obtained from the two-particle correlation $%
g_{ij}(\vec{x}_{i},\vec{x}_{j})$\ and a couple of suitable approximations.

Notice that, in spite of $f_{i}$ being always positive, the function $%
f_{i_{1},...i_{N}}$ is not necessarily so. This fact is exemplified by Eq.(%
\ref{dezenove}), where the relative values of $A$ and $B$ determines if the
pressure is positive or negative.\footnote{%
\, We are not considering phase-transitions effects.}

A remark: the attentive reader will have noticed that in the passage from (%
\ref{vinte dois}) to (\ref{vinte tres}) the dependence on $\vec{v}_{i}$\ has
been supressed. That is because the correlation functions usually depends
only on the position. The $f_{N}$ depend both on position and velocity, but
the sector correspondent to the velocities may be separated, resulting in
non-correlated distribution functions $f_{i}(\vec{v}_{i})$. Figure 1\ shows
schematically the correspondence between the distribution functions and the
clusters for a three-particle system.


\begin{figure}[tbp]
\begin{center}
\centerline{\epsfig{file=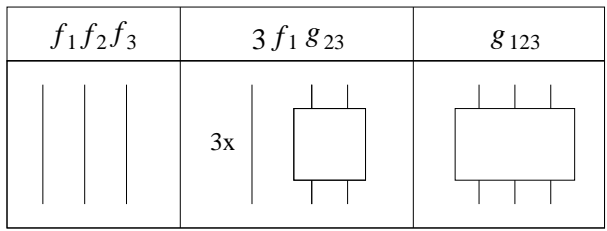,height=3.0cm}}
\end{center}
\caption{{Cluster schematic representation for a three-particle system.}}
\label{fig1}
\end{figure}


The distribution functions for a $N$-particle system are determined, order
by order, by a hierarchical set of equations known as BBGKY -- after
Bogoliubov, Born, Green, Kirkwood and Yvon \cite{BBGKY}. The Boltzmann
equation is itself an approximation of the BBGKY\ system reducing this set
of linear coupled equations to just one non-linear equation. Considering
interaction in this equation will solely modify the one-particle
distribution functions $f_{i}$. The statistical systems modeled by Boltzmann
equation do not exhibit correlation terms in the distribution functions.
Therefore, the thermodynamical quantities, like pressure, do not present
dynamical terms. This leads to the conclusion that the usual approach of
perturbative cosmology does not compute interaction processes as direct
sources of gravitation.

We shall below discuss in which periods of cosmic history interaction should
effectively contribute to gravitation, and how to include it.

\section{On the inclusion of interaction terms \label{sec - inclusao}}

Qualitatively, the cosmic fluid can be separated into three different
sectors: a kinetic part $T$, associated to velocities; a dynamical component
$V$, related to interactions; and a massive contribution $M$ coming from the
rest mass of the constituents. The condition for the interaction being a
relevant source is%
\begin{equation}
V(\Delta t)\geq T(\Delta t)+M(\Delta t),  \label{vinte sete}
\end{equation}%
where $\Delta t$\ indicates a given cosmological period.

As far as interactions are concerned, each period presents its own
characteristics. We will deal with two of then: the
pre-nucleosynthesis cosmological universe (PNU) -- $kT\gtrsim 20$
MeV or red-shift $z\gtrsim 10^{12}$ -- and the recent universe --
red-shift $z\lesssim 20$.

In the PNU, the high temperature of the fluid warrants the existence of a
large variety of particles, such as $\gamma ,\nu ,\pi ,K$, most of then
interacting mutually.\footnote{%
\, The magnitude and type of interaction depends, of course, on the especies
considered.} The answer to the question about the relevance of interactions
for the primeval cosmology comes from an involved analysis of the
interacting fluid as a whole: the interaction terms may be positive or
negative, depending on the nature of the interaction, and global
cancellation may possibly turn up.

Recent data from RHIC (\textit{Relativistic Heavy Ion Collider})
\cite{RHIC} indicates that a fluid at very high temperatures (few
hundreds of MeV) presents a strong interaction between its
constituents, even possibly generating a \textquotedblleft
liquid\textquotedblright\ state for the hadronic matter, the CGC
(\textit{Color Glass Condensate}) \cite{CGC}. Quantum Chromodynamics
suggests that this system is constituted basically by three quarks
($u,d,s$) and gluons. In this energy range -- which corresponds to
the PNU -- the strong interaction between quarks (generically, $q$)
overcome their kinetic and rest energies. We have, so, $V_{q}(\Delta
t_{PNU})\geq T_{q}(\Delta t_{PNU})+M_{q}(\Delta t_{PNU})$. Even
then, the primordial cosmic fluid has others species that should be
considered when applying criterion (\ref{vinte sete}). In section
\ref{sec - exemplo} we exhibit a very simplified model, introducing
interaction terms, to describe the pre-nucleosynthesis cosmological
universe.

Among the various components of the recent universe, one of the most
important for its evolution is non-relativistic matter. Indeed, at the
present-day time $t_{0}$ (red-shift $z=0$), the rest mass of this component $%
M_{NR}$ corresponds to about $30\%$ of the universes' total content \cite%
{Wmap}, and, as we go back in time, it becomes more and more relevant \cite%
{SuNo}. Non-relativistic matter responds to the Newtonian gravitational
interaction $V_{NR}$\ responsible for structure formation and evolution, and
the experimental data indicates the influence of the potential is more
effective as $z$ diminishes (i.e., the structures grow) \cite{HST}. As the
kinetic term $T_{NR}$ is negligible\ compared to $M_{NR}$, the importance of
the gravitational interaction $V_{NR}$ is measured by its direct comparison
with $M_{NR}$. The Newtonian potential is a long range interaction and one
could ask whether it can decisively contribute with dynamical terms that
could influence the present day cosmic behavior. We will not study this
subject in the present work, but only mention three effects controlling its
importance:

\begin{enumerate}
\item As the scale factor $a(t)$\ grows, so does the interaction distance $%
d=a(t)r$ (where $r$\ is the comoving distance), reducing the cosmological
contribution of $V_{NR}$.

\item During a decelerated (accelerated) expansion, the comoving horizon
increases (decreases) consequently increasing (decreasing) the global
effects of $V_{NR}$.

\item The basic constituents of the universe are different at each phase of
the universe evolution, starting with fundamental particles (nucleons,
etc.), passing to hydrogen clouds and then to galaxies and clusters.
\end{enumerate}

Let us return to the analysis of the pre-nucleosynthesis
cosmological period. Next section presents a prescription to include
interaction in that period.


\subsection{Equation of state with interaction terms\label{sec - EOS
interagente}}

The inclusion of interactions via EOS can be done by the ensemble
formalism through the perturbative treatment of real gases. Mayer
and collaborators developed in 1937 the systematic approach of
cluster expansion for a non-relativistic classical (non-quantum)
system ~\cite{mayer}. Just after that, in 1938, Kahn and Uhlenbeck
began the generalization of this method to non-relativistic quantum
statistics \cite{Uhlenbeck,Beth}, and, in 1960, Lee and Yang
improved the treatment to describe, in principle, all the
perturbation orders \cite{LeeYang}. Finally, in 1969, Dashen, Ma and
Bernstein extended this method to a relativistic quantum
system where the interactions are computed through the $S$ matrix~\cite%
{Dashen}. Each one of these treatments apply to a different statistical
system, but all of them were constructed so as to be valid on a plane static
space-time. This prevents their straight application to cosmology, where one
needs to consider the possibility of a curved manifold. This means that the
perturbation methods for modeling real gases are valid only if the
space-section of the universe is plane, and the expansion is of the
quasi-static type. And these requirements, as said in section \ref{sec -
falta}, are fulfilled in the pre-nucleosynthesis universe ($kT\gtrsim 20$ $%
MeV$).

According to the cluster expansion technique, the grand canonical potential
for a one-component fluid is%
\begin{equation}
\Omega (z,T)=\sum\limits_{N=1}^{\infty
}b_{N}~z^{N}~=\sum\limits_{N=1}^{\infty }b_{N}~e^{N\beta \left( \mu
_{NR}+m\right) }~.  \label{vinte oito}
\end{equation}%
The $b_{N}$ are the \textit{cluster integrals} and encapsulate all the
information about the interaction processes. The Appendix is a \textit{resum%
\'{e}} on cluster expansions, with the differences between classical and
quantum systems particularly emphasized.

The first cluster integrals for the non-relativistic classical
system are
\begin{subequations}
\begin{eqnarray}
b_{1} &=&g\frac{e^{-\beta m}}{\lambda ^{3}V}\int d^{3}r_{1}=g\frac{e^{-\beta
m}}{\lambda ^{3}},  \label{vinte nove_a} \\
b_{2} &=&g\frac{e^{-2\beta m}}{2\lambda ^{6}V}\int \int f(\vec{r}_{1},\vec{r}%
_{2})d^{3}r_{1}d^{3}r_{2},  \label{vinte nove_b} \\
b_{3} &=&g\frac{e^{-3\beta m}}{6\lambda ^{9}V}\int \int \int \left[ f(\vec{r}%
_{1},\vec{r}_{2})f(\vec{r}_{1},\vec{r}_{3})f(\vec{r}_{2},\vec{r}_{3})+f(\vec{%
r}_{1},\vec{r}_{2})f(\vec{r}_{1},\vec{r}_{3})+\right.   \notag \\
&&\left. +f(\vec{r}_{1},\vec{r}_{2})f(\vec{r}_{2},\vec{r}_{3})+f(\vec{r}_{1},%
\vec{r}_{3})f(\vec{r}_{2},\vec{r}_{3})\right] d^{3}r_{1}d^{3}r_{2}d^{3}r_{3},
\label{vinte nove_c}
\end{eqnarray}%
where $\lambda $ is the thermal wavelength, $g$ is the degeneracy degree and
\end{subequations}
\begin{equation}
f(\vec{r}_{i},\vec{r}_{j})\equiv e^{-\beta V(\vec{r}_{i},\vec{r}_{j})}-1,
\label{trinta}
\end{equation}%
are the Mayer functions. Notice that in the classical case the
interaction is introduced through the interparticle potential
$V(\vec{r}_{i},\vec{r}_{j}) $. Though the system is nonrelativistic,
the rest mass is already included in the $b_{N}$ through the factors
$e^{-\beta m}$. These are not considered by the traditional texts on
non-relativistic Statistical Mechanics, but they are necessary in
the derivation of coherent cosmological energy density. The pressure
and the numerical density are not affected by these factors.


Dashen, Ma and Berstein \cite{Dashen} have shown that the general form of
the coefficients $b_{N}$\ for a RQS is given in terms of the S-matrix as%
\begin{equation}
b_{N}-b_{N}^{(0)}=\frac{g}{V}\int \frac{e^{-\beta E}}{4\pi i}Tr\left( \hat{A}%
~\hat{S}^{-1}\frac{\overleftrightarrow{\partial }}{\partial E}\hat{S}\right)
_{c_{N}}dE,  \label{trinta um}
\end{equation}%
where $b_{N}^{(0)}$\ is the cluster integral of the non-interacting quantum
theory, $\hat{A}$ is the symmetrization operator, $\hat{S}$\ is the on-shell
S-matrix operator \cite{Dashen Ma}, and $c_{N}$\ stands for all the $N$%
-particle connected diagrams. We are using the definition%
\begin{equation}
\hat{S}^{-1}\frac{\overleftrightarrow{\partial }}{\partial E}\hat{S}=\hat{S}%
^{-1}\frac{\partial \hat{S}}{\partial E}-\frac{\partial \hat{S}^{-1}}{%
\partial E}\hat{S}~.  \label{del arrow}
\end{equation}%
See the Appendix for further details.

Once the grand canonical potential (\ref{vinte oito}) is obtained, it is
straightforward to calculate the pressure (\ref{sete}), the numerical
density (\ref{oito}) and the energy density (\ref{nove})\ as functions of
the temperature and the fugacity:
\begin{subequations}
\begin{gather}
p(z,kT)=kT\sum\limits_{N=1}^{\infty }b_{N}z^{N}~,  \label{trinta dois_a} \\
n(z,kT)=\sum\limits_{N=1}^{\infty }Nb_{N}z^{N}~,  \label{trinta dois_b} \\
\rho (z,kT)=\left( kT\right) ^{2}\sum\limits_{N=1}^{\infty }\frac{\partial
b_{N}}{\partial \left( kT\right) }z^{N}~.  \label{trinta dois_c}
\end{gather}%
These equations are the one-component fluid EOS in parametric form.

An alternative description is obtained if everything is rewritten in terms
of the numerical density after inversion of the $n(z,T)$ series. The result
is the \textit{virial expansion},
\end{subequations}
\begin{subequations}
\begin{gather}
p(n,kT)=kT\sum_{l=1}^{\infty }a_{l}(kT)n^{l}~,  \label{trinta tres_a} \\
\rho (n,kT)=\left( kT\right) ^{2}\sum_{l=1}^{\infty }c_{l}(kT)n^{l}~,
\label{trinta tres_b}
\end{gather}%
where $a_{l}$ and $c_{l}$ are, respectively, the virial coefficients for the
pressure and the energy density. They are completely determined by the $%
b_{N} $. For instance, the first three terms are:
\end{subequations}
\begin{eqnarray}
a_{1} &=&1~,\;\;\;\;\;a_{2}=-\frac{b_{2}}{b_{1}^{2}}~,\;\;\;\;\;a_{3}=\frac{2%
}{b_{1}^{3}}\left( 2\frac{b_{2}^{2}}{b_{1}}-b_{3}\right) ~;
\label{trinta quatro_a} \\
c_{1} &=&\frac{1}{b_{1}}\frac{\partial b_{1}}{\partial \left( kT\right) }=%
\frac{1}{kT}\left( \frac{m}{kT}+\frac{3}{2}\right) ,\;c_{2}=-\frac{\partial
a_{2}}{\partial (kT)}~,\;c_{3}=-\frac{1}{2}\frac{\partial a_{3}}{\partial
(kT)} .  \label{trinta quatro_b}
\end{eqnarray}

The question asking for an answer now is: what is the most suitable set of
EOS for Cosmology, Eqs.(\ref{trinta dois_a}, \ref{trinta dois_b}, \ref%
{trinta dois_c}) or (\ref{trinta tres_a}, \ref{trinta tres_b})? The
following comments will help to determine the preferred set.

The great advantage of writing the EOS in form of series is the possibility
of approximating them by their first terms. Whenever series are involved,
something must be said about their \textit{convergence}. The sets of EOS (%
\ref{trinta dois_a}, \ref{trinta dois_b}, \ref{trinta dois_c}) and (\ref%
{trinta tres_a}, \ref{trinta tres_b}) are equivalent when all the terms of
their series are considered; but that equivalence ceases to exist when the
series are truncated at a given order. The convergence of the sets $\left\{
p(z,T),\rho (z,T)\right\} $ and $\left\{ p(n,T),\rho (n,T)\right\} $ depends
critically on the interaction processes at play and, in consequence, on the
cosmological period under consideration.\

All the relevant particles for the dynamics of the PNU are born through
pair-production at high temperatures (from one tenth to hundreds of MeV).
Each of the species will have~\cite{nar,wein} a nearly vanishing (total)
chemical potential $\mu \simeq 0$ and, therefore, a fugacity close to one, $%
z\simeq 1$. Consequently, $z$ is not a good expansion parameter for a series.

On the other hand, the expansions of $p(n,T)$ and $\rho (n,T)$ describe with
great accuracy rarefied non-relativistic gases. Indeed, using the
Lennard-Jones potential, the correction
\begin{equation}
a_{2}=-\frac{2\pi }{g}\int\limits_{0}^{\infty
}(e^{-V(r_{12})/kT}-1)r_{12}^{2}dr_{12}~,\;\;\;\;\;\;\;r_{12}=\left\vert
\vec{r}_{2}-\vec{r}_{1}\right\vert ~,  \label{trinta cinco}
\end{equation}
leads to an EOS that fits the experimental curves $p\times n$\ for several
gases with very good precision \cite{Hirshefeld}. This is a strong physical
argument in favor of the set $\left\{ p(n,T),\rho (n,T)\right\} $ and
indicates that the associated series are well-defined.

There is another reasoning suggesting the convergence of the series
in $n$: the virial coefficients $a_{l}$ and $c_{l}$ are evaluated
from a subclass of connected diagrams, the \textit{irreducible
diagrams}\emph{\ }\cite{Pat}. In the classical case, the irreducible
diagrams are easily detectable, since they are \textit{multipy
connected}, i.e., each particle connects to at least other two. For
example, coefficient $b_{3}$, given by (\ref{vinte nove_c}), is
constructed from all the four diagrams shown in the right of Figure
2. The coefficients $a_{3}$ and $c_{3}$, however, depend solely on
the last diagram of Figure 2. The first can be written as%
\begin{equation}
a_{3}=-\frac{1}{3g^{2}}\iint f_{12}f_{13}f_{23}d^{3}r_{12}d^{3}r_{13}~.
\label{trinta seis}
\end{equation}%
[Both in (\ref{trinta cinco}) and (\ref{trinta seis}) we are supposing that
the potential depends only on the interparticle distance, $V(\vec{r}_{i},%
\vec{r}_{j})=V(\left\vert \vec{r}_{i}-\vec{r}_{j}\right\vert )$.]
This restriction on the diagrams for the non-relativistic classical
system favors the virial expansion against the series in the
fugacity. It is to be expected that convergence is better achieved
by the virial expansion.


\begin{figure}[tbp]
\begin{center}
\centerline{\epsfig{file=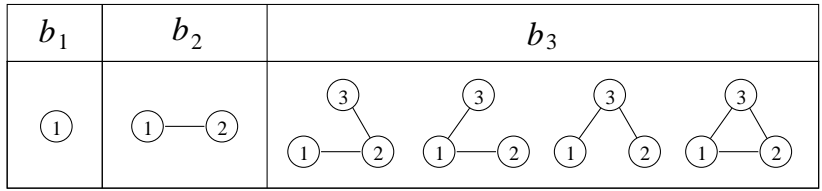,height=2.5cm}}
\end{center}
\caption{{Classical diagrams representing the first three cluster
integrals. Each line corresponds to a Mayer function and each ball,
to a particle.}} \label{fig2}
\end{figure}


In view of these arguments, we choose the set composed by $p(n,T)$
and $\rho (n,T)$ as the most convenient for computing interactions
in the pre-nucleosynthesis cosmological universe. Nevertheless, even
this set presents a validity limit. The perturbative methods do not
apply to dense systems (with high values of $n$) or to systems
submitted to long-range interactions. In such cases the EOS must be
found by other methods.

In next section, we will construct a toy-model in order to exhibit some
possible effects in the evolution of the universe when interaction between
the source components is taken into account. We shall also argue in favor of
the convergence of the series $p(n,T)$ in this particular example, comparing
with the behavior of the expansion $p(z,T)$.


\section{Example of the interaction influence\label{sec - exemplo}}

In our toy-model the primeval interacting fluid is constituted only by
photons and nucleons coming from pair production. The photons $\gamma $ will
be treated in the usual manner as utra-relativistic ideal bosons, and the
nucleons $N$\ will be considered non-relativistic interacting classical
particles.

The interaction processes taken into account are:

\begin{enumerate}
\item Creation and annihilation of the nucleons $N$\ in the thermal bath, $%
\gamma +\gamma \leftrightarrow N+\bar{N}$. This reaction generates the mean
numerical density of nucleons $n_{N}$ and anti-nucleons $n_{\bar{N}}$. In a
purely classical context, this process does not take place. Here, it serves
only as a source of the interacting particles, which are treated classically
as soon as they come into existence.

\item Nucleons affecting nucleons through a (very simplified) nuclear
potential.
\end{enumerate}

The electromagnetic interaction cancels out globally since the numerical
density of nucleons are identical to that of anti-nucleons $\bar{N}$\ (Debye
scenery). Weak interaction is several order of magnitude less effective than
the strong interaction and it is consequently neglected.

For simplicity, we admit that the interactions $NN$, $N\bar{N}$ and $\bar{N}%
\bar{N}$ are described by the same nuclear potential (charge
independence of the strong interaction), and that the internal
degrees of freedom come from spin and isospin. Therefore, a hadronic
part of the cosmic fluid is composed by particles with mass
$m_{N}=m_{\bar{N}}=938.26$ MeV (the proton rest-mass) and degeneracy
$g_{N}=g_{\bar{N}}=4$. The nuclear interaction shall be modeled by a
square-well combined with a hard-core potential, as in Figure 3.

\begin{figure}[tbp]
\begin{center}
\centerline{\epsfig{file=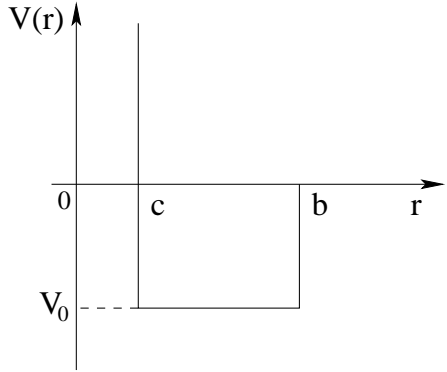,height=5.0cm}}
\end{center}
\caption{{Potential modeling the nuclear interaction.}}
\label{fig3}
\end{figure}
The choice of that nuclear potential is justified by two facts: \textit{(i)}
it presents a behavior similar to those exhibited by some successful
phenomenological nuclear potentials, as those described in Refs.~\cite%
{Reid,Hammada}; and \textit{(ii)} with this simplified potential, it is
possible to calculate analytically the cluster integrals to the third order.

The parameters of the square-well hard-core potential are obtained from (a)
the deuteron binding energy, the proton and deuteron mean-squared radius --
which fix the well's width $\left( b-c\right) =1.3$\ fm and its depth, $%
V_{0}=75.6$\ MeV -- and (b) from nucleons high-energy scattering data --
setting the extension of the hard-core $c=0.4$ $fm$ (cf. Ref. \cite{enge}).

The pressure and the energy density for the PNU are then written in the
form:
\begin{subequations}
\begin{eqnarray}
p(kT,n_{N}) &=&p_{\gamma }(kT)+p_{N}(kT,n_{N})~,  \label{trinta sete_a} \\
\rho (kT,n_{N}) &=&\rho _{\gamma }(kT)+\rho _{N}(kT,n_{N})~,
\label{trinta sete_b}
\end{eqnarray}%
where $p_{N}$ and $\rho _{N}$ are given by (\ref{trinta tres_a}) and (\ref%
{trinta tres_b}) respectively, and \cite{Pat}
\end{subequations}
\begin{equation}
p_{\gamma }=\frac{\rho _{\gamma }}{3}~;\;\;\;\;\;\rho _{\gamma }(kT)=\frac{%
\pi ^{2}(kT)^{4}}{15}~.  \label{trinta oito}
\end{equation}%
Truncating the nucleons EOS in the third order (approximation valid for $%
n_{N}$ not too large), explicit expressions for $p(T,n_{N})$ and $\rho
(T,n_{N})$ result:
\begin{subequations}
\begin{eqnarray}
p(kT,n_{N}) &\simeq &\frac{\pi ^{2}(kT)^{4}}{45}+\left( kT\right) \left[
n_{N}+a_{2}n_{N}^{2}+a_{3}n_{N}^{3}\right] ~,  \label{trinta nove_a} \\
\rho (kT,n_{N}) &\simeq &{\textstyle{\frac{\pi ^{2}(kT)^{4}}{15}}}+{%
\textstyle{\left( m_{N}+\frac{3}{2}kT\right) }}n_{N}+\left( kT\right) ^{2}%
\left[ c_{2}n_{N}^{2}+c_{3}n_{N}^{3}\right] ,  \label{trinta nove_b}
\end{eqnarray}%
with
\end{subequations}
\begin{equation}
n_{N}(kT)\simeq g_{N}\frac{e^{-\beta m_{N}}}{\lambda _{N}^{3}}%
+2b_{2}+3b_{3}~.  \label{quarenta}
\end{equation}%
Recall that $\mu _{N}\simeq 0$ in the pre-nucleosynthesis
cosmological period and, in consequence, $z_{N}=1$.
Coefficients $a_{2}$ and $a_{3}$ are found analytically with the help of (%
\ref{trinta cinco}) and (\ref{trinta seis}), while $b_{2}$, $b_{3}$ and $%
c_{2}$, $c_{3}$\ are derived directly from (\ref{trinta quatro_a}) and (\ref%
{trinta quatro_b}). Performing all calculations and inserting $n_{N}(kT)$ in
(\ref{trinta nove_a}) and (\ref{trinta nove_b}), we finally obtain the EOS $%
p(kT)$ and $\rho (kT)$ for the PNU. The result is presented in a graphic
form -- see Figure 4. 
\begin{figure}[tbp]
\epsfig{file=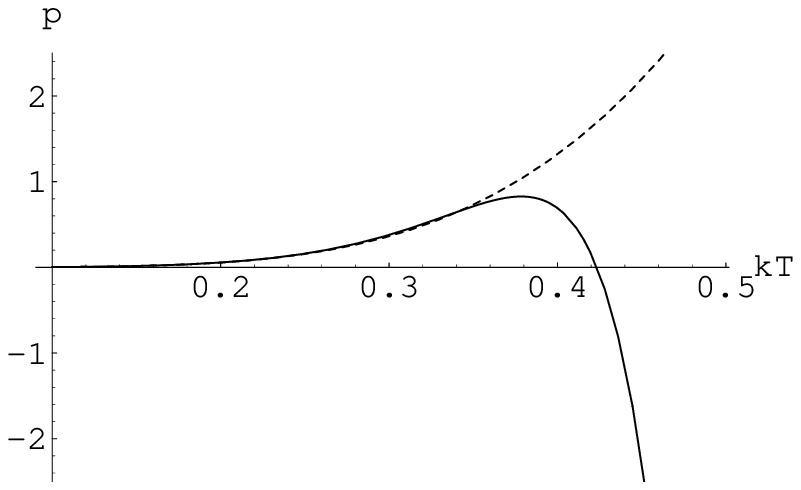,height=3.6cm} %
\epsfig{file=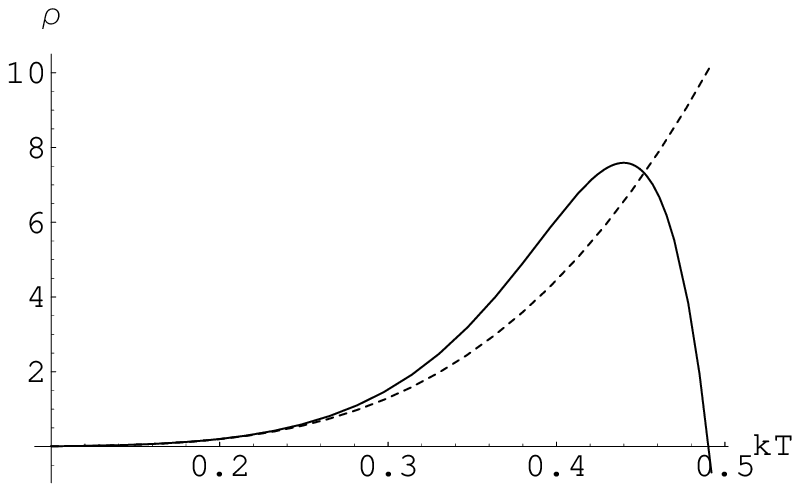,height=3.6cm} 
\caption{{Graphics of the pressure $p(kT)$ and energy density $\protect\rho %
(kT)$, both measured in GeV/fm$^{3}$, as functions of $kT$ given in GeV.
Full lines represent $p$ and $\protect\rho $ of the proposed model (with
interaction). For sake of comparison, the dotted curves show $p$ and $%
\protect\rho $ in the ideal case (without interaction).}}
\label{fig4}
\end{figure}
For $kT$ $\lesssim $ $0.3$ GeV, the proposed model is qualitatively
identical to the ideal case. This is an expected feature since, until $0.3$
GeV, the numerical density $n_{N}(kT)$ is too small to cause any relevant
interaction effect. From this energy value\ on, the deviation from the ideal
case emerges, first in the curve for $\rho $ and then in the plot of $p$. As
energy increases, the two functions tend to decrease and, eventually, $p$
and $\rho $ become negative. This peculiar characteristic is due to the
action of the interaction terms in (\ref{trinta nove_a}) and (\ref{trinta
nove_b}). They dominate at high-energies: the attractive part of the nuclear
potential (square-well) make $a_{2}$ and $c_{2}$ the most important terms of
the truncated expansion.

The effect of the nuclear interaction in the PNU is better seen by observing
the equation of state in its modern--cosmology familiar form,
\begin{equation}
w(kT)\equiv \frac{p(kT)}{\rho (kT)} \,\, .  \label{quarenta um}
\end{equation}

\begin{figure}[tbp]
\begin{center}
\centerline{\epsfig{file=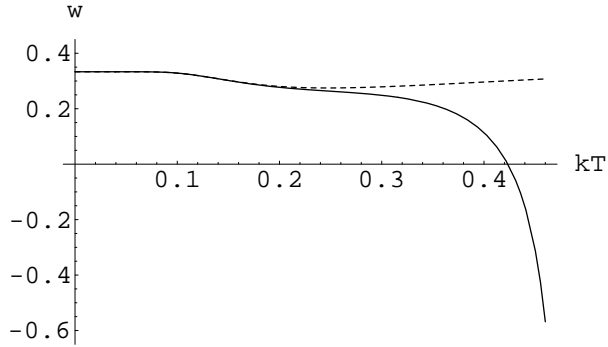,height=5.0cm}}
\end{center}
\caption{{Parametric equation $w(kT)$ as a function of $kT$ (in GeV). The
full curve represents the proposed model (with interaction). The dotted line
corresponds to the ideal (without interaction) cosmological EOS.}}
\label{fig5}
\end{figure}


The behavior of $w(kT)$\ is shown in Figure 5. For $kT<0.1$ GeV the function
$w(kT)\simeq 1/3$, implying that the EOS is that typical of a
radiation--dominated universe. The ideal terms reduce slightly this value of
$w$ as the energy increases. But, for energy values greater than $0.3$ GeV,
the interaction processes become relevant, reducing abruptly the value of $%
w(kT)$ and deviating its behavior from the ideal case.

The dynamical (interaction) terms modify the cosmological EOS and,
consequently, change the form of the primordial expansion. In
particular, our toy-model produces an accelarated expansion which
naturally evolves to a decelerated radiation-like expansion, which
is necessary for the nucleosynthesis. Nevertheless, we would not
dare to say that this simplistic model represents realistically the
PNU dynamics. This model is apt only  to show the importance of
considering interactions within the cosmic fluid as an important
factor in the determination of the early universe's evolution.

Another argument favoring the better convergence of the pressure series $%
p(kT,n_{N})$ (\ref{trinta nove_a}) in comparison to that for the
energy-density series $p(kT,z_{N})$ (\ref{trinta dois_a}) follows as a
by-product. It is possible to compare the second interaction terms ($a_{2}$
or $b_{2}$) with the third ones ($a_{3}$ or $b_{3}$), by defining the
functions%
\begin{equation}
F_{n}(kT)\equiv 1-\left\vert \frac{a_{3}n_{N}^{3}}{a_{2}n_{N}^{2}}%
\right\vert ~;\;\;\;\;\;F_{z}(kT)\equiv 1-\left\vert \frac{b_{3}z_{N}^{3}}{%
b_{2}z_{N}^{2}}\right\vert ~.  \label{quarenta tres}
\end{equation}
The more are $F_{n}(kT)$ and $F_{z}(kT)$ close to $1$, the more the
third interactions terms are irrelevant compared to the second
terms, and the larger is the possibility of convergence of
$p(kT,n_{N})$ and $p(kT,z_{N})$. Notice, however, that this analysis
does not proves convergence: it uses only the first terms of the
series. $F_{n}(kT)$ and $F_{z}(kT)$ only help us to find an argument
favoring the good behavior of the EOS. Graphics of $F_{n} $ and
$F_{z}$ as functions of $kT$ are showed in Figure 6. The conclusion
is that the series in terms of $n_{N}$ -- namely $p(kT,n_{N})$ --
has a better chance to converge than the series in $z_{N}$. This
corroborates the choice of section \ref{sec - EOS interagente},
where we have chosen the EOS set given in terms of the numerical
density as the suitable EOS for the pre-nucleosynthesis cosmological
universe.
\begin{figure}[tbp]
\begin{center}
\centerline{\epsfig{file=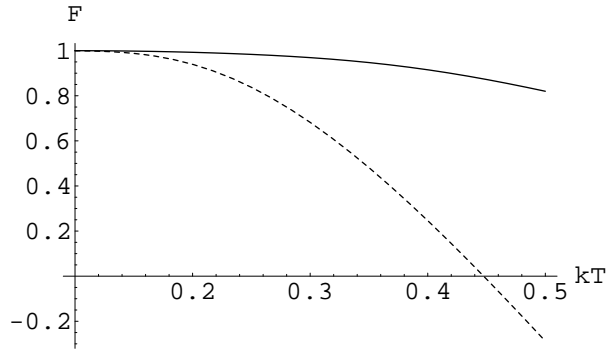,height=5.0cm}}
\end{center}
\caption{{The functions $F_{n}(kT)$ (full line) and $F_{z}(kT)$ (dashed
line). $kT$ is measured in GeV.}}
\label{fig6}
\end{figure}


\section{Final remarks \label{sec - Dis. Final}}


This work highlights the absence of dynamical terms as direct source
of curvature in the usual cosmological models. The detailed
examination of this fact is performed in section \ref{sec - falta}.
The analysis is trivial for the equilibrium approach, but, even in
the non-equilibrium case, no mention is made of the possible
relevance of interacting components for the universe evolution. Some
exceptions are Refs. \cite{Ber,Ruben1,Bascos}.

A method for including these dynamical effects in the pre-nucleosynthesis
universe has been proposed, through the corrections of the EOS based in the
cluster expansion technique of Statistical Mechanics. The approach is rather
general and allows the treatment of the cosmic fluid as a classical or
quantum system (relativistic or not). In principle, the method can be
applied to other periods of cosmic history, provided the conditions for
thermodynamical equilibrium and series convergence are respected.\footnote{%
\, Such conditions are required only for the interacting part of the cosmic
fluid.}

A toy-model has been presented which illustrates the deep consequences which
interaction between constituents can have for cosmic evolution. Even if
overmuch simplified, the model points the way toward more realistic
approaches, based on fundamental physics. More accurate models for the
pre-nucleosynthesis universe would include other particles than just photons
and nucleon--anti-nucleons pairs: at least pions, kaons, electrons and
neutrinos should be included. Besides, the nuclear interaction should be
treated in a more complete manner than just a square-well hard-core
potential. More realistic EOS are found in  studies \cite%
{Pelaez,Greiner,Kapusta1} concerning interacting hadrons.

We cannot affirm categorically that the inclusion of dynamical terms can
describe properly the primordial acceleration of the universe (as suggested
by our toy-model), but the results presented here are indicative that these
terms should not be simply ignored, as usually done. Maybe the inflationary
era and the present-day acceleration are just consequences of neglected
interaction terms.

\appendix

\section{Appendix: Cluster expansions}

The ensemble approach to Statistical Physics is able to include interactions
perturbatively, via the cluster expansion formalism. The method is developed
in the grand canonical ensemble, whose partition function, written for a
single component fluid, is:
\begin{equation}
\Xi (z,V,T)=\sum\limits_{N=0}^{\infty }Q_{N}(V,T)z^{N}  \label{A1}
\end{equation}%
with $z$ is given by (\ref{z}) and
\begin{equation}
Q_{N}(V,T)=\frac{1}{N!}\int\limits_{\Omega }W_{N}^{X}(\vec{r}_{1},...,\vec{r}%
_{N})d^{3}r_{1}...d^{3}r_{N},  \label{A2}
\end{equation}%
where tag $X$ indicates the nature of the system under
consideration: $X=C$ denotes a non-relativistic classical system and
$X=Q$, a quantum system (QS). $W_{N}^{X}$ is the probability density
for a $N$-particle system,
\begin{subequations}
\begin{eqnarray}
W_{N}^{C}(\vec{r}_{1},...,\vec{r}_{N}) &\equiv &\left( \frac{e^{-\beta m}}{%
\lambda ^{3}}\right) ^{N}\exp \left[ -\beta \sum\limits_{i<j=1}^{N}V(\vec{r}%
_{i},\vec{r}_{j})\right] ~,  \label{A3} \\
W_{N}^{Q}(\vec{r}_{1},...,\vec{r}_{N}) &\equiv &N!\left\langle \vec{r}%
_{1},...,\vec{r}_{N}\right\vert \hat{A}e^{-\beta \hat{H}_{N}}\left\vert \vec{%
r}_{1},...,\vec{r}_{N}\right\rangle ~,  \label{A4}
\end{eqnarray}%
where $V(\vec{r}_{i},\vec{r}_{j})$ is the potential between the particles $i$
and $j$; $\hat{A}$ is the symmetrization operator and $\hat{H}_{N}$ the $N$%
-particle Hamiltonian operator.


The thermodynamical quantities are associated to the grand canonical
potential $\Omega $ which, just like the partition function, can be
expressed as a series in terms of the fugacity $z$,
\end{subequations}
\begin{equation}
\Omega (z,T)=\frac{1}{V}\ln \Xi (z,V,T)\equiv \sum\limits_{N=1}^{\infty
}b_{N}z^{N}~,  \label{A5}
\end{equation}%
with%
\begin{equation}
b_{N}=\frac{g}{N!V}\int\limits_{\Omega }U_{N}^{X}(\vec{r}_{1},...,\vec{r}%
_{N})d^{3}r_{1}...d^{3}r_{N}~.  \label{A6}
\end{equation}%
The $U_{N}^{X}$\ are the \textit{Ursell functions} and $g$\ counts the
degeneracy coming from internal degrees of freedom.\footnote{%
\, How to count the degenerate states depends criticaly on the type of
system. For instance, when dealing with a quantum system, the counting must
respect the symmetry condition of the complete wave-function.} The
coefficients $b_{N}$ are the \textit{cluster integrals}.\

Using Eqs.(\ref{A1}), (\ref{A2}), (\ref{A5}) and (\ref{A6}) and expanding $%
\ln \Xi $ in terms of the fugacity, one writes $U_{N}^{X}$ in terms of $%
W_{N}^{X}$. Examples:
\begin{gather}
U_{1}^{X}(\vec{r}_{1}) =W_{1}^{X}(\vec{r}_{1})~,  \notag \\
U_{2}^{X}(\vec{r}_{1},\vec{r}_{2}) =W_{2}^{X}(\vec{r}_{1},\vec{r}
_{2})-W_{1}^{X}(\vec{r}_{1})W_{1}^{X}(\vec{r}_{2})~,  \notag \\
U_{3}^{X}(\vec{r}_{1},\vec{r}_{2},\vec{r}_{3}) = \qquad \qquad \qquad \qquad
\qquad \qquad \qquad \qquad \qquad \qquad \qquad \qquad \qquad  \notag \\
W_{3}^{X}(\vec{r}_{1},\vec{r}_{2},\vec{r}_{3})-3W_{2}^{X}(\vec{r}_{1},\vec{r}%
_{2})W_{1}^{X}(\vec{r}_{3})+2W_{1}^{X}(\vec{r}_{1})W_{1}^{X}(\vec{r}%
_{2})W_{1}^{X}(\vec{r}_{3}).  \label{A7}
\end{gather}

In principle, all the possible states of a statistical system can be
decomposed in generic diagrams accounting for the several correlation
processes (interactions or quantum exchange effects). The $Q_{N}$,
determined from the probability density functions, are obtained as the sum
of all distinct diagrams of the $N$ particles. The cluster integrals,
determined from the Ursell functions, are obtained as the sum of all
connected diagrams,
\begin{equation}
b_{N}=\frac{g}{V}\sum \left( \text{connected\textit{\ diagrams }of }N\text{
particles}\right) ~.  \label{A8}
\end{equation}%
In a $N$-particle connected diagram all the $N$ particles are linked
directly or indirectly. The links include all kinds of correlation.

The arrangement of the several distinct diagrams in connected diagrams is
complicated and it will not be carried out here. The interested reader may
consult Refs. \cite{Pat} for classical systems, and \cite{Dashen} for the
quantum case.



\subsection{Non-relativistic classical system}

According to (\ref{A3}):%
\begin{eqnarray}
U_{1}^{C}(\vec{r}_{1}) &=&W_{1}^{C}(\vec{r}_{1})=\frac{e^{-\beta m}}{\lambda
^{3}}~,  \label{A9a} \\
U_{2}^{C}(\vec{r}_{1},\vec{r}_{2}) &=&\left( \frac{e^{-\beta m}}{\lambda ^{3}%
}\right) ^{2}f(\vec{r}_{1},\vec{r}_{2})~,  \label{A9b}
\end{eqnarray}%
where%
\begin{equation}
f(\vec{r}_{i},\vec{r}_{j})\equiv e^{-\beta V(\vec{r}_{i},\vec{r}_{j})}-1
\label{A10}
\end{equation}%
is the \textit{Mayer function}.

Also from (\ref{A3}), it is possible to show that the
non-relativistic classical cluster functions can be decomposed in
products of Mayer functions \cite{Hirshefeld}. For
instance: 
\begin{equation*}
U_{3}^{C}(\vec{r}_{1},\vec{r}_{2},\vec{r}_{3})={\textstyle{\left( \frac{%
e^{-\beta m}}{\lambda ^{3}}\right) ^{3}}}\left( e^{-\beta V(\vec{r}_{1},\vec{%
r}_{2})}e^{-\beta V(\vec{r}_{1},\vec{r}_{3})}e^{-\beta V(\vec{r}_{2},\vec{r}%
_{3})}-3e^{-\beta V(\vec{r}_{1},\vec{r}_{2})}+2\right) .
\end{equation*}%
\begin{multline}
U_{3}^{C}=\left( \frac{e^{-\beta m}}{\lambda ^{3}}\right) ^{3}\left[ \left(
e^{-\beta V(\vec{r}_{1},\vec{r}_{2})}-1\right) \left( e^{-\beta V(\vec{r}%
_{1},\vec{r}_{3})}-1\right) \left( e^{-\beta V(\vec{r}_{2},\vec{r}%
_{3})}-1\right) +\right.   \notag \\
+e^{-\beta V(\vec{r}_{1},\vec{r}_{2})}e^{-\beta V(\vec{r}_{1},\vec{r}%
_{3})}+e^{-\beta V(\vec{r}_{1},\vec{r}_{2})}e^{-\beta V(\vec{r}_{2},\vec{r}%
_{3})}+e^{-\beta V(\vec{r}_{1},\vec{r}_{3})}e^{-\beta V(\vec{r}_{2},\vec{r}%
_{3})}+ \\
\left. -\,2e^{-\beta V(\vec{r}_{1},\vec{r}_{2})}-2e^{-\beta V(\vec{r}_{1},%
\vec{r}_{3})}-2e^{-\beta V(\vec{r}_{1},\vec{r}_{2})}+3\right] ~.
\end{multline}%
\begin{gather}
U_{3}^{C}(\vec{r}_{1},\vec{r}_{2},\vec{r}_{3})=\left( \frac{e^{-\beta m}}{%
\lambda ^{3}}\right) ^{3}\left[ f(\vec{r}_{1},\vec{r}_{2})f(\vec{r}_{1},\vec{%
r}_{3})f(\vec{r}_{2},\vec{r}_{3})+f(\vec{r}_{1},\vec{r}_{2})f(\vec{r}_{1},%
\vec{r}_{3})\right.   \notag \\
\left. +f(\vec{r}_{1},\vec{r}_{2})f(\vec{r}_{2},\vec{r}_{3})+f(\vec{r}_{1},%
\vec{r}_{3})f(\vec{r}_{2},\vec{r}_{3})\right] ~.  \label{A11}
\end{gather}%
Such a decomposition can be obtained for all orders. The Mayer functions
will always be the fundamental entities in the determination of the $%
U_{N}^{C}(\vec{r}_{1},...,\vec{r}_{N})$, and lead to a
representation of the cluster integrals in terms of connected
diagrams~---~the first three are shown in Figure 2.

The number of connected diagrams grows very fast as $N$ increases: there is
only one diagram if $N=2$, $4$ diagrams for $N=3$, and $38$ for$N=4$.

\subsection{Relativistic quantum system}

Relativistic quantum system is much more complicated than the
non-relativistic classical system by several reasons:

\begin{itemize}
\item In a relativistic system an interaction cannot be represented by a
potential. In fact, the determination of an interaction Hamiltonian $%
H^{\prime }\equiv H-H_{0}$, the total Hamiltonian $H$ minus the free
Hamiltonian $H_{0}$, is a difficult task.

\item There are quantum correlation effects which are mixed to the dynamical
(interaction) terms.

\item There are discrete bound states.
\end{itemize}

Despite these difficulties, there exists a general formalism giving the
coefficients $b_{N}$ in terms of the S-matrix \cite{Dashen}.

Analogously to (\ref{A4}), we define the operator $\hat{U}_{N}$ such as%
\begin{equation}
U_{N}^{Q}(\vec{r}_{1},...,\vec{r}_{N})\equiv N!\left\langle \vec{r}_{1},...,%
\vec{r}_{N}\right\vert \hat{U}_{N}\left\vert \vec{r}_{1},...,\vec{r}%
_{N}\right\rangle ~.  \label{A12}
\end{equation}%
Then, using (\ref{A6}),%
\begin{equation}
b_{N}=\frac{g}{V}\int\limits_{\Omega }\left\langle \vec{r}_{1},...,\vec{r}%
_{N}\right\vert \hat{U}_{N}\left\vert \vec{r}_{1},...,\vec{r}%
_{N}\right\rangle d^{3}r_{1}...d^{3}r_{N}~,  \label{A13}
\end{equation}%
or through a Fourier transform,
\begin{equation}
b_{N}=\frac{g}{V}\int\limits_{\Omega _{k}}\left\langle \vec{k}_{1},...,\vec{k%
}_{N}\right\vert \hat{U}_{N}\left\vert \vec{k}_{1},...,\vec{k}%
_{N}\right\rangle d^{3}k_{1}...d^{3}k_{N}~.  \label{A14}
\end{equation}%
The general form of the $b_{N}$ is%
\begin{equation}
b_{N} = \frac{g}{V} \, Tr\hat{U}_{N}~.  \label{A15}
\end{equation}%
Equation (\ref{A13}) is the realization of this trace in coordinate
representation, and (\ref{A14}) is its realization in the momentum
representation.

In order to obtain the $b_{N}$\ in terms of the S-matrix, it is necessary to
separate the statistical effects from the dynamical effects, because only
the latter influence the S-matrix. The traditional method to accomplish this
is to construct%
\begin{equation}
b_{N}^{(0)}=\frac{g}{V}\,Tr\hat{U}_{N}^{(0)}\ ,  \label{A16}
\end{equation}%
where index $\left( 0\right) $ denotes a free system (without interaction). $%
\hat{U}_{N}^{(0)}$ is related to a free system, it is responsible only for
statistical contribution. Therefore, it is sufficient to subtract (\ref{A16}%
) from (\ref{A15}) to retain the pure dynamical terms. This gives%
\begin{equation}
b_{N}-b_{N}^{(0)}=\frac{g}{V}Tr\left( \hat{U}_{N}-\hat{U}_{N}^{(0)}\right) ~.
\label{A17}
\end{equation}%
Thus, the grand canonical potential (\ref{A5}) is:%
\begin{equation}
\Omega -\Omega _{0}=\sum\limits_{N=1}^{\infty }\left(
b_{N}-b_{N}^{(0)}\right) z^{N}~,  \label{A18}
\end{equation}%
where $\Omega _{0}$ is the ideal grand canonical potential (\ref{seis}).
Dashen, Ma and Bernstein \cite{Dashen} showed that (\ref{A17}) can be
written in terms of the S-matrix as:%
\begin{equation}
Tr\left( \hat{U}_{N}-\hat{U}_{N}^{(0)}\right) =\int \frac{e^{-\beta E}}{4\pi
i}Tr\left( \hat{A}.\hat{S}^{-1}\frac{\overleftrightarrow{\partial }}{%
\partial E}\hat{S}\right) _{c_{N}}dE~,  \label{A19}
\end{equation}%
where $\hat{A}$ is the symmetrization operator, $\hat{S}$\ is the on-shell
S-matrix operator \cite{Dashen Ma}, and $c_{N}$\ indicate that only the
connected $N$-particle diagrams are to be considered.

Taking (\ref{A19}) into (\ref{A17}) leads to the general expression
(\ref{trinta um}) for the cluster integrals $b_{N}$. That form of
$b_{N}$ includes not only the
 scattering states, but also the (bound state) composed
particles (for more details see \cite{Dashen}).

Let us exhibit the explicit expressions of (\ref{A19}) for one-particle and
two-particle systems.

\begin{enumerate}
\item $N=1$\textit{:}%
\begin{equation}
Tr\left( \hat{U}_{1}-\hat{U}_{1}^{(0)}\right) =Tr\left( \hat{W}_{1}\right)
-Tr\left( \hat{W}_{1}^{(0)}\right) =Tr\hat{A}e^{-\beta \hat{H}_{1}}-Tr\hat{A}%
e^{-\beta \hat{H}_{1}^{(0)}}=0~.  \label{A20}
\end{equation}%
The last equality results from $\hat{H}_{1}=\hat{H}_{1}^{(0)}$ (free
particle).

\item $N=2$:
\begin{subequations}
\begin{equation*}
Tr\left( \hat{U}_{2}-\hat{U}_{2}^{(0)}\right) =Tr\,\hat{W}_{2}-Tr\left( \hat{%
W}_{1}\hat{W}_{1}\right) -\left( Tr\,\hat{W}_{2}^{(0)}-Tr\left( \hat{W}%
_{1}^{(0)}\hat{W}_{1}^{(0)}\right) \right) .
\end{equation*}%
\end{subequations}
\begin{equation*}
Tr\left( \hat{U}_{2}-\hat{U}_{2}^{(0)}\right) =Tr\left( \hat{W}_{2}\right)
-Tr\left( \hat{W}_{2}^{(0)}\right) ~.
\end{equation*}%
\begin{align}
Tr\left( \hat{U}_{2}-\hat{U}_{2}^{(0)}\right) & =Tr\hat{A}e^{-\beta \hat{H}%
_{2}}-Tr\hat{A}e^{-\beta \hat{H}_{2}^{(0)}}  \notag \\
& =\int \frac{e^{-\beta E}}{4\pi i}Tr\left( \hat{A}\hat{S}_{2}^{-1}\frac{%
\overleftrightarrow{\partial }}{\partial E}\hat{S}_{2}\right) dE,
\label{A21}
\end{align}%
where $\hat{H}_{2}$ is the two-particle Hamiltonian operator and $\hat{S}_{2}
$ is the S-matrix operator associated to $\hat{H}_{2}$. It is worth to
remark that (\ref{A21}) can be rewritten in terms of measurable \emph{\ }%
\textit{phase-shifts}, if we choose the angular momentum representation.
\end{enumerate}


\end{document}